\documentstyle[bezier,epsfig,12pt]{article}

\begin{document}
\leftline{\hglue 10cm TMUP-HEL-9706}
\leftline{\hglue 10cm hep-ph/9706546}
\baselineskip=19pt
\vskip 0.7in
\begin{center}
{\large{\bf THREE FLAVOR NEUTRINO OSCILLATION ANALYSIS OF}}
{\large{\bf THE KAMIOKANDE ATMOSPHERIC NEUTRINO DATA}}
\end{center}
\vskip 0.4in
\begin{center}
Osamu Yasuda\footnote{Email: yasuda@phys.metro-u.ac.jp}
\vskip 0.2in
{\it Department of Physics, Tokyo Metropolitan University}

{\it 1-1 Minami-Osawa Hachioji, Tokyo 192-03, Japan}
\end{center}

\vskip .7in
\centerline{ {\bf Abstract} }

Using the published Kamiokande binned data for sub-GeV and
multi-GeV atmospheric neutrinos, we have searched for the optimum set of
three flavor neutrino oscillation parameters within the
constraints of reactor experiments.  It is found that $\chi^2$ is
minimized for ($\Delta m_{21}^2$, $\Delta m_{32}^2$,
$\theta_{12}$, $\theta_{13}$, $\theta_{23}$) = (7.4$\times
10^{-1}~$eV$^2$, 2.6$\times 10^{-2}~$eV$^2$, $2^\circ$, $3^\circ$,
$45^\circ$) and (2.6$\times 10^{-2}~$eV$^2$, 7.4$\times
10^{-1}~$eV$^2$, $0^\circ$, $87^\circ$,
$46^\circ$) with
$\chi_{\rm min}^2$ = 87.8 (76\%CL).  The sets of parameters which are
suggested by the two flavor analysis turn out to be close to the
minimum (0.2$\sigma$).

\newpage

There has been much interest in atmospheric neutrinos
\cite{kamioka1,kamioka2,imb,frejus,nusex,soudan2}, 
which might give evidence for neutrino oscillations.  While
NUSEX \cite{nusex} and Frejus \cite{frejus} have reported consistency
between the data and the predictions of the atmospheric neutrino flux
\cite{flux,hkkm,hkm},
Kamiokande, IMB and Soudan-2 have reported a discrepancy.  In
particular, the Kamiokande group claimed that their data suggests that the
mass squared difference is of order 10$^{-2}$eV$^2$
\cite{kamioka2}.  The reason that they have obtained a narrow
region for the mass squared difference is because they have used both
the binned data in the sub-GeV and the multi-GeV energy regions.
Namely, while the multi-GeV data show remarkable zenith angle
dependence, the sub-GeV data have little zenith angle dependence.
This indicates that the mass squared difference relevant to the
neutrino oscillations in the atmospheric neutrinos cannot be much
smaller than or much larger than 10$^{-2}$eV$^2$.  The momentum
spectra of the sub-GeV atmospheric neutrinos (Fig.1 in
\cite{kamioka1}) also supports this argument.

People have studied neutrino oscillations among three flavors
\cite{kp}, and it has been shown recently \cite{3nu} that one can
easily get strong constraints for the mass squared differences and the
mixing angles if one assumes a mass hierarchy.  The original analysis of
the atmospheric neutrino data by the Kamiokande group
\cite{kamioka1,kamioka2} was based on the framework of neutrino
oscillation between two flavors and it is important to see what
happens if we analyze the data in the three flavor framework.  Much
work has been done on the analysis of atmospheric neutrinos
\cite{atmth,zenith1,zenith2,yasuda} and among those that
discuss quantitatively the binned
data of the multi-GeV neutrinos are
\cite{zenith1,zenith2,yasuda}.  However, these works did not
take the sub-GeV binned data of Kamiokande \cite{kamioka1} into
account, and the allowed regions for the relevant mass squared
difference obtained in there are wider than those in \cite{kamioka2}.
In this paper\footnote{This paper supersedes \cite{yasuda}.} we will
perform a three flavor analysis of the published data
\cite{kamioka1,kamioka2} of the Kamiokande atmospheric neutrino
experiment, along with the data of the reactor neutrino experiments
\cite{bugey,kras}.  We will
make full use of both the sub-GeV and the multi-GeV binned data as
well as the data of reactor neutrino experiments \cite{bugey,kras},
and as we will see below, we have strong constraints on the mass
squared differences and some of the mixing angles.  We will take the
smearing effects into consideration, and evaluate the number of events
by summing over the energy and the zenith angle of neutrinos, to
reproduce the original analysis by the Kamiokande group as much as
possible.  Throughout this paper we will restrict our discussions only
to the data by the Kamiokande group\cite{kamioka1,kamioka2}, not only
because the Monte Carlo result for the neutrino energy spectrum is
available only in Ref.
\cite{kamioka1,kamioka2}, but also because this is the only data which gives
both the upper and the lower bound on the mass squared difference of
neutrinos.

We start with the Dirac equation for three flavors of neutrinos
with mass in matter \cite{msw}
\begin{eqnarray}
i\frac{d}{dx} \Psi (x) =
\left[ U {\rm diag}\left(0, \Delta m_{21}^2/2E, \Delta m_{31}^2/2E \right)
U^{-1} + {\rm diag}\left( A(x),0,0 \right)
\right] \Psi (x),
\label{eqn:dirac}
\end{eqnarray}
where we have taken the ultra-relativistic limit and have subtracted
a term proportional to the unit matrix,
$\Delta m_{ij}^2\equiv m_i^2-m_j^2$ is the mass squared difference
of the neutrinos with energy $E$,
$\Psi(x) \equiv (\nu_e(x), \nu_\mu(x), \nu_\tau(x))^T$ is
the wave function of the neutrinos in the flavor basis,
and $A(x) \equiv \sqrt{2} G_F N_e(x)$~stands for the effect
due to the charged current interactions between $\nu_e$
and electrons in matter \cite{msw}.  Here,
\begin{eqnarray}
U\equiv\left( U_{\alpha j} \right)
\equiv\left(
\begin{array}{ccc}
c_{12}c_{13} & s_{12}c_{13} &   s_{13}\\
-s_{12}c_{23}-c_{12}s_{23}s_{13} & c_{12}c_{23}-s_{12}s_{23}s_{13}
& s_{23}c_{13}\\
s_{12}s_{23}-c_{12}c_{23}s_{13} & -c_{12}s_{23}-s_{12}c_{23}s_{13}
& c_{23}c_{13}
\end{array}
\right)
\end{eqnarray}
with $c_{ij}\equiv\cos\theta_{ij},~s_{ij}\equiv\sin\theta_{ij}$
($\alpha=e,\mu,\tau$; $i,j=1,2,3$)
is the orthogonal mixing matrix of neutrinos.  For simplicity, we will not
discuss the CP violating phase of the mixing matrix here.
\footnote{Even if we include the CP violating phase $\delta$
of the mixing matrix, the effect of $\delta$ always appears in the
combination of $c_{13}\sin 2\theta_{13}\sin\delta$.  $\sin
2\theta_{13}$ has to be small because of the constraints from the
reactor experiments.}.  Without loss of
generality we can assume $m_1^2<m_2^2<m_3^2$, so that we have $\Delta
m_{31}^2>\max(\Delta m_{21}^2,\Delta m_{32}^2)$, and we will use
$\Delta m_{21}^2,\Delta m_{32}^2>0$ as two independent parameters.

The number of the expected charged leptons $\ell_\alpha$~
( $\ell_\alpha$ = $e$ or $\mu$ ) with energy $q$ from a scattering
$\nu_\alpha N
\rightarrow \ell_\alpha N'$ $(\alpha=e,\mu)$
is given by
\begin{eqnarray}
\displaystyle
N(\ell_\alpha)
&=& n_T\sum_{\beta=e,\mu}
\int_0^\infty dE
\int^{q_{\rm max}}_0 dq
\int_{-1}^1 d\cos\Theta
\int_{-1}^1 d\cos\theta
\int_0^{2\pi} d\varphi~\epsilon_\alpha (q)\nonumber\\
&\times&
{d^3F_\beta (E,\theta) \over dE~d\cos\theta~d\varphi}
\cdot{ d^2\sigma_\alpha (E,q) \over dq~d\cos\psi }
\cdot{d\cos\psi \over d\cos\theta}
{\ }P(\nu_\beta\rightarrow\nu_\alpha; E, \theta)
\label{eqn:n}
\end{eqnarray}
Here $d^3F_\beta /dEd\cos\theta d\varphi$ is the flux of atmospheric
neutrinos $\nu_\beta$ with energy $E$ from the zenith angle $\theta$,
$n_T$ is
the effective number of target nucleons, $\epsilon_\alpha (q)$ is
the detection
efficiency function for charged leptons $\ell_\alpha$,
$d\sigma_\alpha/dqd\cos\psi$ is the differential cross section
of the interaction $\nu_\alpha N \rightarrow \ell_\alpha N'$
($\alpha$ = $e$ or $\mu$), and $\Theta$ is the zenith angle
of the direction from which the charged lepton $\ell_\alpha$ comes
(See Fig. 1).
\vglue 0.5truecm
(Insert Fig.1 here.)
\vglue 0.5truecm
\noindent
$P(\nu_\beta\rightarrow\nu_\alpha; E, \theta)$ is the probability
of $\nu_\beta\rightarrow\nu_\alpha$ transitions with energy $E$ after
traveling a distance
$\displaystyle L=\sqrt{(R+h)^2-R^2\sin^2\theta}-R\cos\theta$,
where $R$ is the radius of the Earth, $h\sim$15Km is the altitude
at which atmospheric neutrinos are produced.

As for the analysis of the sub-GeV data, almost all the information
which is necessary to get the right hand side of (\ref{eqn:n}) is
available in the published references.  We have used the differential
cross section $d\sigma_\alpha / dqd\cos\psi$ in \cite{gnc}, and the
detection efficiency function $\epsilon_\alpha (q)$ is given in
\cite{kajita1}.  The flux of atmospheric neutrinos
$d^3F_\beta /dEd\cos\theta d\varphi$ without the geomagnetic effects
is given in \cite{hkkm} but we have used the flux which has been
obtained with the geomagnetic effects \cite{hkm}.  There are two
important types of binned data in \cite{kamioka1} from which we derive
the dependence of neutrino oscillations on the energy
and the path length of neutrinos: Fig.1 (a) and (b) in \cite{kamioka1}
give the binned data with respect to energy of the outgoing charged
leptons, while Fig.3 in \cite{kamioka1} gives the binned data (the
ratio $(\mu/e)_{\rm data}/(\mu/e)_{\rm MC}$) with respect to the
zenith angle.  These binned data are expressed as
\begin{eqnarray}
\displaystyle
Y_j^\alpha
&=& (1+\alpha)~n_T\sum_{\beta=e,\mu}C_\beta
\int_{\cos\Theta_j}^{\cos\Theta_{j+1}} d\cos\Theta
\int_0^\infty dE
\int^{q_{\rm max}}_0 dq
\int_{-1}^1 d\cos\theta
\int_0^{2\pi} d\varphi\nonumber\\
&{ }&\times\epsilon_\alpha (q)\cdot
{d^3F_\beta (E,\theta) \over dE~d\cos\theta~d\varphi}
\cdot{ d^2\sigma_\alpha (E,q) \over dq~d\cos\psi }
\cdot{d\cos\psi \over d\cos\Theta}
{\ }P(\nu_\beta\rightarrow\nu_\alpha; E, \theta)\\
Y_a^\alpha
&=& (1+\alpha)~n_T\sum_{\beta=e,\mu}C_\beta
\int^{q_a}_{q_{a+1}} dq
\int_0^\infty dE
\int_{-1}^1 d\cos\Theta
\int_{-1}^1 d\cos\theta
\int_0^{2\pi} d\varphi\nonumber\\
&{ }&\times\epsilon_\alpha (q)\cdot
{d^3F_\beta (E,\theta) \over dE~d\cos\theta~d\varphi}
\cdot{ d^2\sigma_\alpha (E,q) \over dq~d\cos\psi }
\cdot{d\cos\psi \over d\cos\Theta}
{\ }P(\nu_\beta\rightarrow\nu_\alpha; E, \theta),
\end{eqnarray}
respectively, where $(1+\alpha)$ is a factor for the uncertainty in
the absolute normalization, $C_\mu\equiv1+\beta/2$,
$C_e\equiv1-\beta/2$ for the uncertainty in the relative normalization
of the $\nu_\mu$and $\nu_e$ flux, and
$\cos\Theta_j\equiv (2j-7)/5$ ($1 \le j \le 5$), $q_a\equiv (a+1)/10$ GeV
($1 \le a \le 10$).

To reproduce the analysis of the multi-GeV data by the Kamiokande group,
one needs the detection efficiency function $\epsilon_\alpha (q)$,
which is not given in \cite{kamioka2}.
However, the quantity
\begin{eqnarray}
g_{\alpha} (E)&=&
 n_T\int^{q_{\rm max}}_0 dq~\epsilon (q)
\int_{-1}^1d\cos\Theta
\int_{-1}^1d\cos\theta\int_0^{2\pi}d\varphi\nonumber\\
&{ }&\times
{d^3F_\alpha (E,\theta) \over dE~d\cos\theta}
\cdot{ d^2\sigma_\alpha (E,q) \over dq~d\cos\psi}
\cdot{d\cos\psi \over d\cos\Theta}
\label{eqn:flux0}
\end{eqnarray}
is given in Fig.2 (d)--(f) in Ref. \cite{kamioka2}.
The zenith angle dependence $n_\beta (E, \Theta)$ of the atmospheric
neutrino flux for various neutrino energy $E$ has been given
in Ref. \cite{hkkm} in detail.  Here we multiply the quantity
$g_{\alpha} (E)$
by the zenith angle dependence $n_\alpha (E, \Theta)$
in Ref. \cite{hkkm,hkm} with suitable
normalization
$N'\equiv 1/\int_0^\pi d\cos\Theta ~n_\alpha
(E,\Theta)$, and insert the Gaussian factor
with $1\sigma\equiv 17^\circ$ to simulate a smearing effect of the
detecting neutrinos.  Thus we adopt the following quantity
for the multi-GeV analysis.
\begin{eqnarray}
X^\alpha_j
&\equiv&NN'(1+\alpha)\sum_{\beta=e,\mu}C_\beta
\int_{\Theta_j}^{\Theta_{j+1}} d\cos\Theta \int_0^\infty dE
\int\int_D d\cos\theta d\varphi\nonumber\\
&\times&{g_{\beta} (E) n_{\beta} (E,\Theta)
\exp\left(-\tan^2\psi/2 / \tan^2\psi_0/2
\right) \over (1+\cos\psi)
\sqrt{\sin^2\Theta-\sin^2\theta\sin^2\varphi}}
P(\nu_\beta\rightarrow\nu_\alpha;E,\theta),
\label{eqn:multi}
\end{eqnarray}
where $D$ is the region of $(\theta,\varphi)$ in which the argument
of $\sqrt{\sin^2\Theta-\sin^2\theta\sin^2\varphi}$ becomes
positive, $N\equiv 2 /\sqrt{\pi}\tan(\psi_0/2)$ is the normalization
factor such that
\begin{eqnarray}
N\int_{-1}^1 d\cos\psi
{\exp(-\tan^2\psi/2/\tan^2\psi_0/2) \over
(1+\cos\psi)^{3/2}(1-\cos\psi)^{1/2}}=1,
\end{eqnarray}
and we put $\psi_0=17^\circ$.  The relation between $\psi$ and $\Theta$
\begin{eqnarray}
\cos\psi={\cos\theta\cos\Theta-\sin\theta\cos\varphi
\sqrt{\sin^2\Theta-\sin^2\theta\sin^2\varphi} \over
1-\sin^2\theta\sin^2\varphi}
\end{eqnarray}
can be obtained by spherical trigonometry, and the measure
$d\cos\psi/d\cos\Theta$ gives factors in the numerator in
(\ref{eqn:multi}).  (\ref{eqn:multi}) is not exactly the same quantity
as the one in the original analysis
\cite{kamioka2}, but this is almost the best which can
be done with the published data in \cite{kamioka2}.

Several groups \cite{flux,hkkm} have given predictions on the
flux of atmospheric neutrinos but they differ from one another in the
magnitudes, and the Kamiokande group assumed that the errors of the
overall normalization $1+\alpha$ and the relative normalization $1+\beta/2$
are $\sigma_\alpha$=30\% and $\sigma_\beta$=12\%, respectively.
Thus we define the total $\chi^2$ as \cite{bc}
\begin{eqnarray}
\chi^2={\alpha^2 \over \sigma_\alpha^2}
+{\beta^2 \over \sigma_\beta^2}
+\chi_{\rm sub-GeV}^2+\chi_{\rm multi-GeV}^2
+\chi_{\rm Bugey}^2+\chi_{\rm Krasnoyarsk}^2,
\label{eqn:chi}
\end{eqnarray}
where
\begin{eqnarray}
\displaystyle\chi_{\rm sub-GeV}^2&=&
2\sum_{\alpha=e,\mu}\sum_{a=1}^{10}
\left( Y_a^\alpha-y_a^\alpha-y_a^\alpha
\ln{Y_a^\alpha \over y_a^\alpha} \right)
+\sum_{j=1}^5{1 \over \sigma_j^2}
\left( {Y_j^\mu/y_j^\mu \over Y_j^e/y_j^e}
-r_j \right)^2,\nonumber\\
\\
\displaystyle\chi_{\rm multi-GeV}^2&=&
2\sum_{\alpha=e,\mu}\sum_{j=1}^5
\left( X_j^\alpha-x_j^\alpha-x_j^\alpha
\ln{X_j^\alpha \over x_j^\alpha} \right),
\end{eqnarray}
$\chi_{\rm Bugey}^2$ is given by (9) in \cite{bugey}, and $\chi_{\rm
Krasnoyarsk}^2$ is defined in terms of the eight data and the three
parameters $N_1,N_3,N_b$ defined in \cite{kras}.
Here $x_j^\alpha,y_j^\alpha~(\alpha=e,\mu;1\le j \le 5)$
are the multi-GeV and the sub-GeV data for each zenith angle
$\cos\Theta_j<\cos\Theta<\cos\Theta_{j+1}$, respectively,
and $y_a^\alpha~(\alpha=e,\mu;1\le a \le 10)$ are the sub-GeV data of the
charged leptons with the energy $q_a<q<q_{a+1}$.  It is understood that
$\alpha^2 / \sigma_\alpha^2~+~\beta^2 / \sigma_\beta^2 ~+~\chi_{\rm
sub-GeV}^2+\chi_{\rm multi-GeV}^2$ is the optimum value with respect
to $\alpha, \beta$, $\chi_{\rm Bugey}^2$ with respect to the
parameters $A, b, a_j (j=1,2,3)$ defined in \cite{bugey}, and
$\chi_{\rm Krasnoyarsk}^2$ with respect to the parameters
$N_1,N_3,N_b$.  The theoretical predictions
$X_j^\alpha,Y_j^\alpha~(1\le j \le 5)$, $Y_a^\alpha~(1\le a \le 10)$,
$(\alpha=e,\mu)$ depend on five free parameters ($\Delta m_{21}^2$,
$\Delta m_{32}^2$; $\theta_{12}$, $\theta_{13}$, $\theta_{23}$), where
$\Delta m_{ij}\equiv m_i^2-m_j^2$, so (\ref{eqn:chi}) is expected to
obey a $\chi^2$ distribution with 103$-$5=98 degrees of freedom
(10$\times$2+5=25 for sub-GeV, 5$\times$2=10 for multi-GeV, 60 for
Bugey, and 8 for Krasnoyarsk).  The number of degrees of freedom in
the present analysis for the sub-GeV and the multi-GeV data is
smaller than in the original one by the Kamiokande group
(10$\times$11=110 for sub-GeV, 5$\times$8+5=85 for multi-GeV).

The value of $\chi^2$ is affected to some extent by the presence of
matter, and it is necessary to take into consideration the
contribution of the second term in (\ref{eqn:dirac}).  In the case of
the multi-GeV analysis we have solved (\ref{eqn:dirac}) numerically
for each $E$ (0.9 GeV $\le$ $E$ $\le$ 100 GeV) and evaluated the
number of events for a given range of the zenith angle.  As for the
sub-GeV case, however, it turned out that the second term in
(\ref{eqn:dirac}) is negligible for the entire
energy region of the sub-GeV neutrinos with $\Delta m_{ij}^2$ under
consideration ($A(x)\ll\Delta m_{ij}^2 / E $), so we have ignored the
matter contribution in the sub-GeV analysis.  We have also
taken into account the effects of the particle misidentification error
and the backgrounds ($Y^e\rightarrow Y^e+0.04Y^\mu$, $Y^\mu\rightarrow
0.96Y^\mu$ for sub-GeV \cite{kamioka1}, $X^e\rightarrow
X^e+0.08X^\mu$, $X^\mu\rightarrow 0.92X^\mu$ for multi-GeV \cite{kajita2}
\footnote{The 4\% uncertainty (out of 8\%) of the multi-GeV data is due
to the $\mu$-like neutral current events \cite{nakahata}.}).  We have
almost reproduced the Monte Carlo results by the Kamiokande group,
such as the energy spectrum of the sub-GeV (Fig. 1 in \cite{kamioka1})
the zenith angle distributions of the sub-GeV (Fig. 2 in
\cite{kamioka1}) and the multi-GeV data (Fig. 3 in \cite{kamioka2}).

We have meshed each $\Delta m_{ij}^2$ region into 18 points
($\Delta m_{ij}^2 =10^{(2\ell-24)/5},~(0\le \ell \le 17)$)
and each $\theta_{ij}$ region into 4 points
$\theta_{ij}=\ell\pi / 6,~(0\le \ell \le 3)$, and
the evaluated the value of $\chi^2$.  Furthermore, using the
grid-search and the
gradient-search methods described in Ref. \cite{br}, we have
found that $\chi^2$ has the minimum value for
\begin{eqnarray}
&{ }&(\Delta m_{21}^2,\Delta m_{32}^2)=
(7.4\times 10^{-1}{\rm eV}^2,2.6\times 10^{-2}{\rm eV}^2)\nonumber\\
&{ }&(\theta_{12},\theta_{13},\theta_{23})=
(2^\circ,3^\circ,45^\circ)\nonumber\\
&{ }&(\alpha,\beta)=(3.1\times10^{-1},-6.5\times10^{-2})
\end{eqnarray}
with $\chi^2_{\rm min}=87.8$
and
\begin{eqnarray}
&{ }&(\Delta m_{21}^2,\Delta m_{32}^2)=
(2.6\times 10^{-2}{\rm eV}^2,7.1\times 10^{-2}{\rm eV}^2)\nonumber\\
&{ }&(\theta_{12},\theta_{13},\theta_{23})=
(0^\circ,87^\circ,46^\circ)\nonumber\\
&{ }&(\alpha,\beta)=(3.1\times10^{-1},-6.9\times10^{-2})
\end{eqnarray}
with $\chi^2_{\rm min}=87.8.$

The number of degrees of freedom of our analysis is 98, so the value
of the reduced chi square is 0.9, which corresponds to 76 \%
confidence level.  This suggests that our fit in the present analysis
is good.  The 68 \% CL, 90 \% CL allowed regions of the parameters
$\Delta m_{21}^2, \Delta m_{32}^2$ with unconstrained $\theta_{12},
\theta_{13}, \theta_{23}$ are given by $\chi^2 \le \chi^2_{\rm
min}+5.9$, $\chi^2 \le \chi^2_{\rm min}+9.2$ respectively, and are
given in Fig. 2, where the asterisks stand for the sets of the
parameters for $\chi^2_{\rm min}$.  The reason that we have obtained a
rather narrow region for ($\Delta m_{21}^2, \Delta m_{32}^2$) is
because we have taken into consideration the multi-GeV binned
data, the sub-GeV binned data and the reactor data together:
The significant zenith angle dependence of the multi-GeV data
gives certain lower and upper
bounds on $\Delta m_{ij}^2$ and the little zenith angle dependence of
the sub-GeV data gives more constraint on the lower bound.  In general
$\chi_{\rm multi-GeV}^2$ prefers a large value of $\sin^22\theta_{13}$
(if $\Delta m_{21}^2 < \Delta m_{32}^2$) or a large value of
$4|U_{e1}|^2(1-|U_{e1}|^2)$ (if $\Delta m_{21}^2 > \Delta m_{32}^2$),
but the reactor experiments exclude a region with large value of
either factor if the larger $\Delta m_{ij}^2$ is of order
10$^{-2}$eV$^2$.

\vglue 0.5truecm
(Insert Fig.2 here.)
\vglue 0.5truecm

In Fig. 3 we have plotted $\chi^2$ as a function of $\theta_{13}$,
putting $\Delta m_{21}^2=0$ with $\theta_{23}$ unconstrained.  This is
a situation which is realized by the two flavor analysis of the solar
neutrino problem \cite{msw,solar}; Irrespective of whether we consider
the vacuum solution ($\Delta m_{21}^2\sim{\cal O}(10^{-11}$eV$^2$)) or
the MSW solution ($\Delta m_{21}^2\sim{\cal O}(10^{-5}$eV$^2$)) for
the solar neutrino, the mass squared difference $\Delta m_{21}^2$ is
negligible compared to the contribution of the matter effect $A(x)$ in
(\ref{eqn:dirac}) and the other mass squared difference $\Delta
m_{32}^2$, which should be at least of order 10$^{-2}$eV$^2$ to
account for the zenith angle dependence of the multi-GeV
data.  In this case $\theta_{12}$ does not appear in $\chi^2$.  In
particular, for $\theta_{13}=0$ we have $\chi^2-\chi^2_{\rm min}=2.2$,
so we conclude that this set of parameters falls within $0.2\sigma$
for all three types of the solar neutrino solutions\cite{msw,solar}
\footnote{In fact any set of the parameters which satisfies
$\Delta m_{21}^2\ll\Delta m_{32}^2<\Delta m_{31}^2$ and
$\theta_{13}=0$ falls within $0.2\sigma$.}.

\vglue 0.5truecm
(Insert Fig.3 here.)
\vglue 0.5truecm

In this paper we have analyzed the sub-GeV and the multi-GeV
atmospheric neutrino data by the Kamiokande group based on the
framework of three flavor neutrino oscillations along with
constraints from the reactor experiments, and have shown that
at least one of $\Delta m_{ij}^2$ should be of order
10$^{-2}$eV$^2$ at 90 \% confidence level.  We have
also shown that the popular set of parameters ( $(\Delta
m_{21}^2,\sin^2 2\theta_{12})=(\Delta m^2,\sin^2 2\theta)_\odot$,
$(\Delta m_{32}^2,\sin^2 2\theta_{23})=
(\Delta m^2,\sin^2 2\theta)_{\rm atm}$, $\theta_{13}=0$)
fall within 0.2$\sigma$,
which is very close to the best fit.  The
minimum value of $\chi^2$ is 87.8 for 98 degrees of freedom, and the fit
based on the hypothesis of neutrino oscillations is
good.  If we combine
the present results with the solar neutrino experiments, then we get even
stronger constraints, which will be reported somewhere \cite{yasuda2}.

\vskip 0.2in
\noindent
{\Large{\bf Acknowledgement}}
\vskip 0.1in

The author would like to thank H. Minakata for discussions, M. Honda
for giving him the unpublished results of the atmospheric neutrino
flux calculations, T. Kajita for useful communications, and
S.D.H. Hsu for careful reading of the manuscript.  This
research was supported in part by a Grant-in-Aid for Scientific Research
of the Ministry of Education, Science and Culture, \#09045036.

\newpage
\noindent
{\Large{\bf Figures}}

\begin{description}
\item[Fig.1] The parametrization of angles
in the interaction $\nu_\alpha+N\rightarrow \ell_\alpha+N'$.

\item[Fig.2] The allowed region for $\Delta m_{21}^2, \Delta m_{32}^2$
with $\theta_{ij}$ unconstrained.  The solid and dashed lines
stand for the regions at 90 \% and 68 \% confidence level, respectively.
The asterisks stand for the sets of the parameters for the best fit.

\item[Fig.3] $\chi^2$ as a function of $\theta_{13}$
with $\Delta m_{21}^2=0$ and $\theta_{23}$ unconstrained.
$\chi^2$ is independent of $\theta_{12}$ in this limit.
\end{description}
\newpage
\pagestyle{empty}
\unitlength=.2mm
\begin{center}
\begin{picture}(100,100)(0,450)
\put(-50,400){\makebox(100,100){\Huge{$\nu_\alpha$}}}
\put(70, 270){\makebox(100,100){\Huge{$\psi$}}}
\put(220, 220){\makebox(100,100){\Huge{$\ell_\alpha$}}}
\put(-90, 270){\makebox(100,100){\Huge{$\theta$}}}
\put(-150,230){\makebox(100,100){\Huge{$e_z$}}}
\put(10, 130){\makebox(100,100){\Huge{$\Theta$}}}
\put(-30, -130){\makebox(100,100){\Huge{$\varphi$}}}
\put(-30, -430){\makebox(100,100){\Huge{\bf Fig.1}}}
\put(0,0){\vector(0,1){400}}
\put(0,0){\line(1,0){350}}
\put(0,0){\vector(1,1){235}}
\put(0,0){\vector(-1,3){81}}
\put(0,0){\line(2,-1){185}}
\put(185,185){\line(0,-1){277}}
\put(0,0){\line(-3,-2){200}}
\bezier{150}(0,300)(100,300)(185,185)
\bezier{150}(0,300)(-50, 260)(-60.,185)
\bezier{150}(185,185)(20,255)(-60.,185)
\bezier{150}(-39,-25)(0,-45)(50.,-25)
\end{picture}
\end{center}
\newpage
\epsfig{file=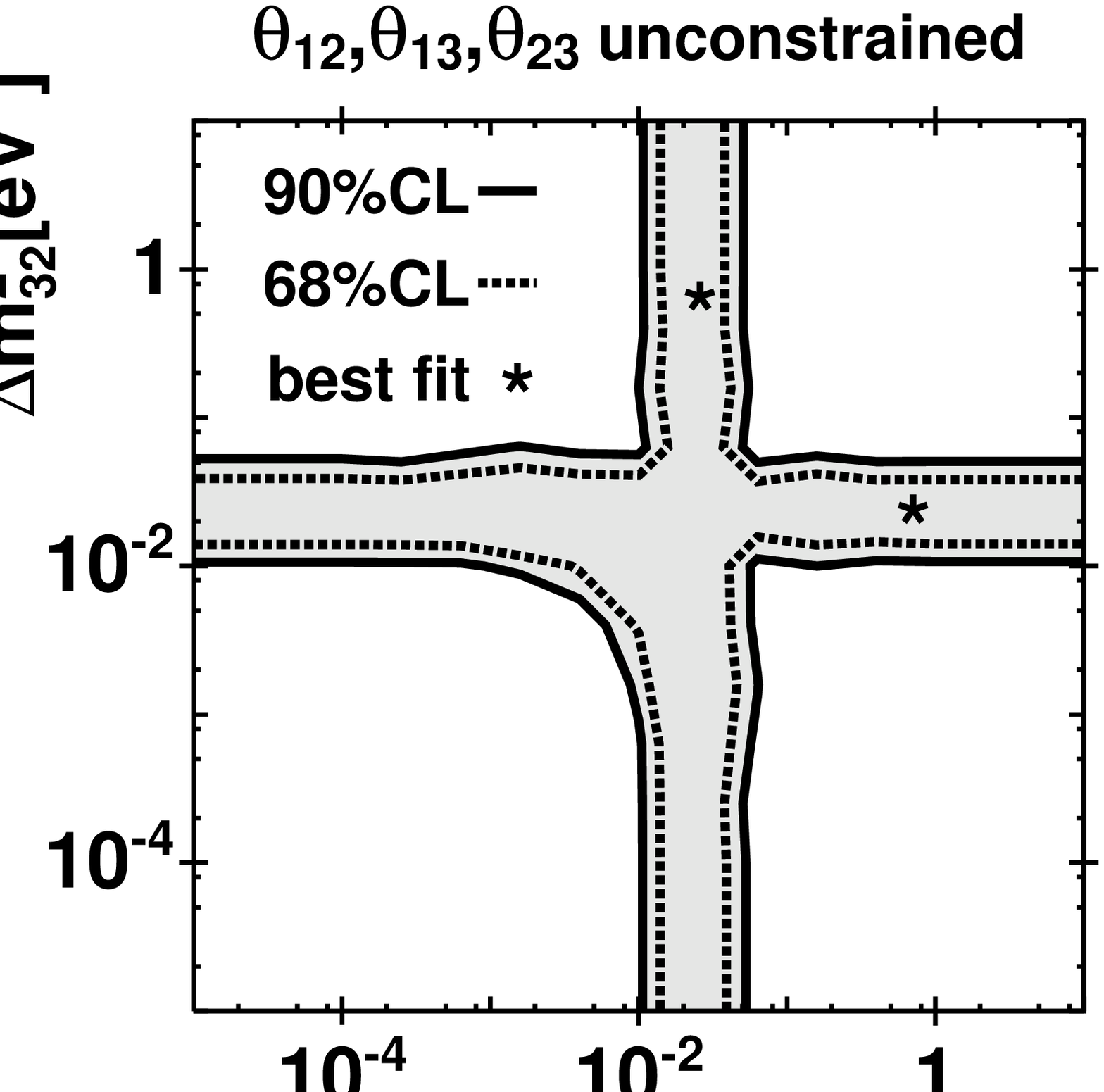,width=15cm}
\newpage
\epsfig{file=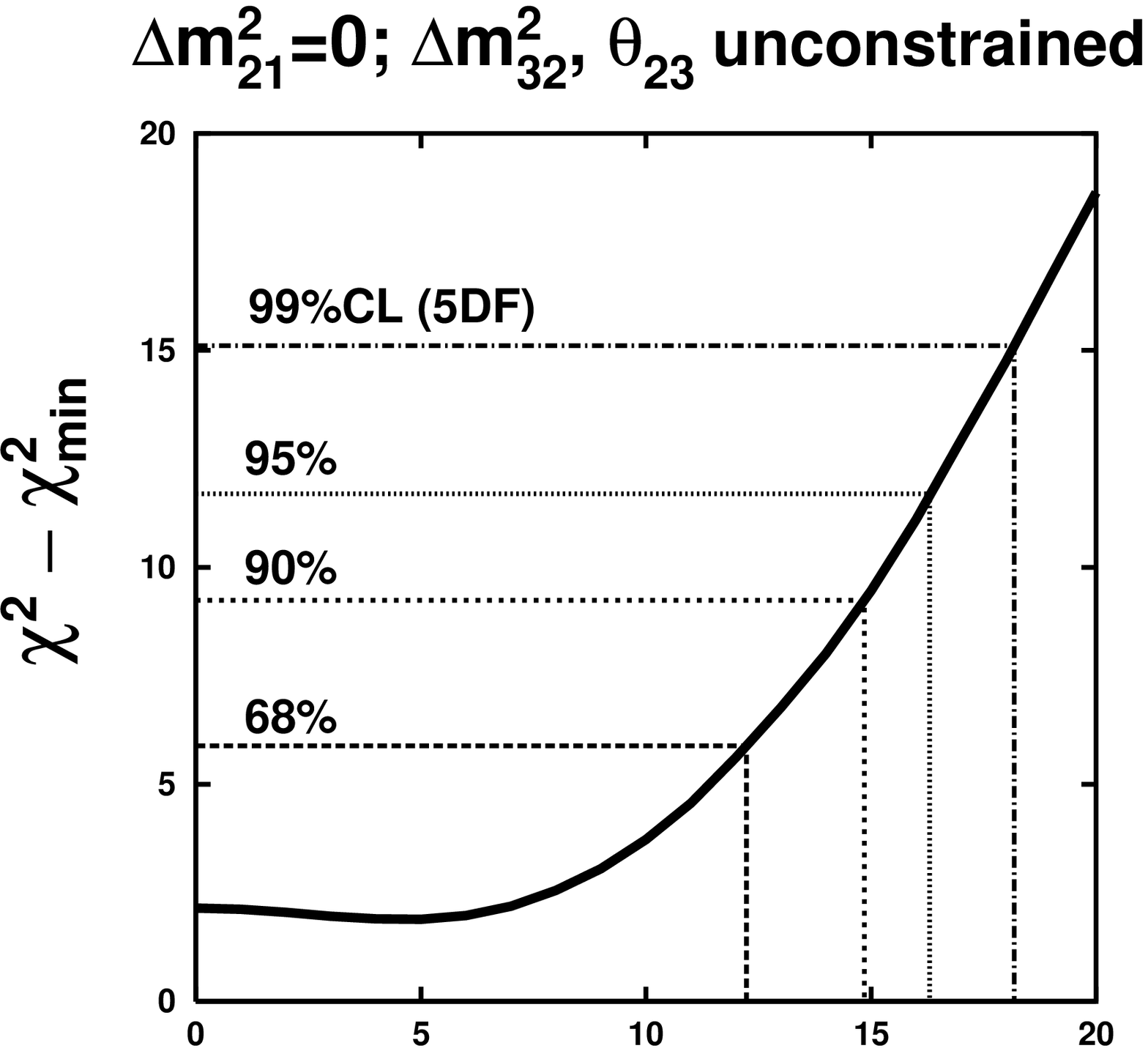,width=15cm}


\newpage
\begin{thebibliography}{99}
\bibitem{kamioka1}Kamiokande Collaboration, K.S. Hirata et al.,  Phys. Lett. {\bf B205} (1988) 416;
{\it ibid.} {\bf B280} (1992) 146. 
\bibitem{kamioka2}
Kamiokande Collaboration, Y. Fukuda et al., Phys. Lett. {\bf B335} (1994) 237. 
\bibitem{imb}
IMB Collaboration, D. Casper et al., Phys. Rev. Lett. {\bf 66} (1989) 2561;
R. Becker-Szendy et al., Phys. Rev. {\bf D46} (1989) 3720.
\bibitem{nusex}
NUSEX Collaboration, M. Aglietta et al., Europhys. Lett. {\bf 8} (1989) 611.
\bibitem{frejus}
Frejus Collaboration, Ch. Berger et al.,  Phys. Lett. {\bf B227} (1989) 489;
{\it ibid.} {\bf B245} (1990) 305; K. Daum et al, Z. Phys. {\bf C66}
(1995) 417. 
\bibitem{soudan2}
Soudan 2 Collaboration, M. Goodman et al., Nucl. Phys. {\bf B}
(Proc. Suppl.) {\bf 38} (1995) 337;
W.W.M., Allison et. al., Phys. Lett. {\bf B391} (1997) 491.
\bibitem{flux}
L.V. Volkova, Sov. J. Nucl. Phys. {\bf 31} (1980) 784;
T.K. Gaisser, T. Stanev S.A. Bludman and H. Lee, Phys. Rev. Lett.
{\bf 51} (1983) 223;
A. Dar, Phys. Rev. Lett.
{\bf 51} (1983) 227;
K. Mitsui, Y. Minorikawa and H. Komori, Nuovo Cim. {\bf C9} (1986) 995;
E.V. Bugaev and V.A. Naumov, Sov. J. Nucl. Phys. {\bf 45} (1987) 857;
T.K. Gaisser, T. Stanev and G. Bar, Phys. Rev. {\bf D38} (1988) 85;
A.V. Butkevich, L.G. Dedenko and I.M. Zheleznykh, Sov. J. Nucl. Phys.
{\bf 50} (1989) 90;
M. Honda, K. Kasahara, K. Hidaka and S. Midorikawa, Phys. Lett. {\bf B248},
193 (1990);
H. Lee and Y. S. Koh, Nuovo Cim. {\bf B105} (1990) 883;
M. Kawasaki and S. Mizuta, Phys. Rev. {\bf D43} (1991) 2900;
P. Lipari, Astropart. Phys. {\bf 1} (1993) 195.
D.H. Perkins, Astropart. Phys. {\bf 2} (1994) 249;
V. Agrawal, T.K. Gaisser, P. Lipari and T. Stanev,
Phys. Rev. {\bf D53} (1996) 1314.
\bibitem{hkkm}
M. Honda, T. Kajita, S. Midorikawa, and K. Kasahara,
Phys. Rev. {\bf D52} (1995) 4985.
\bibitem{hkm}
M. Honda, K. Kasahara, and S. Midorikawa, private communication.
\bibitem{kp}
See T.K. Kuo and J. Pantaleone, Rev. Mod. Phys. {\bf 61} (1989) 937
and references therein.
\bibitem{3nu}
G.L. Fogli, E. Lisi, D. Montanino, Phys. Rev. {\bf D49} (1994) 3626;
G. Conforto, Nucl. Phys. {\bf B} (Proc. Suppl.) {\bf 38} (1995) 308;
C. Giunti, C.W. Kim and J.D. Kim, Phys. Lett. {\bf B352} (1995) 357;
H. Minakata, Phys. Lett. {\bf B356} (1995) 61;
Phys. Rev. {\bf D52} (1995) 6630;
S.M. Bilenky, A. Bottino, C. Giunti, C.W. Kim,
Phys. Lett. {\bf B356} (1995) 273; Phys. Rev. {\bf D54} (1996) 1881;
S.M. Bilenky, C. Giunti, C.W. Kim and S.T. Petcov,
Phys. Rev. {\bf D54} (1996) 4432;
K.S. Babu, J.C. Pati and F. Wilczek, Phys. Lett. {\bf B359} (1995) 351,
{\bf B364} (1995) 251 (E);
M. Narayan, M.V.N. Murthy, G. Rajasekaran, S. Uma Sankar,
Phys. Rev. {\bf D53} (1996) 2809;
G.L. Fogli, E. Lisi and G. Scioscia, Phys. Rev. {\bf D52} (1995) 5334;
S. Goswami, K. Kar and A. Raychaudhuri,
Int. J. Mod. Phys. {\bf A12} (1997) 781.
\bibitem{atmth}
J.G. Learned, S. Pakvasa, and T.J. Weiler, 
Phys. Lett. {\bf B207}, 79 (1988);
{\it ibid.} {\bf B298}, 149 (1993);
V. Barger and K. Whisnant, 
Phys. Lett. B {\bf 209}, 365 (1988);
K. Hidaka, M. Honda and S. Midorikawa,
Phys. Rev. Lett. {\bf 61} (1988) 1537;
S. Midorikawa, M. Honda and K. Kasahara,
Phys. Rev. {\bf D44} (1991) R3379;
J. Pantaleone, Phys. Rev. {\bf D43} (1991) 641;
{\it ibid.} {\bf D49} (1994) 2152;
X. Shi and D.N. Schramm, Phys. Lett. {\bf B283} (1992) 305;
T.K. Kuo and J. Pantaleone, Phys. Rev. {\bf D47} (1993) 4059;
A. Acker, A.B. Balantekin and F. Loreti, Phys. Rev. {\bf D49} (1994) 328.
\bibitem{zenith1}
G.L. Fogli, E. Lisi, Phys. Rev. {\bf D52} (1995) 2775;
D. Saltzberg, Phys. Lett. {\bf B355} (1995) 499.
\bibitem{zenith2}
S.M. Bilenky, C. Giunti and C.W. Kim, Astropart. Phys.
{\bf 4} (1996) 241;
G.L. Fogli, E. Lisi, D. Montanino and G. Scioscia,
Phys. Rev. {\bf D55} (1997) 4385;
M. Narayan, G. Rajasekaran and S. Uma Sankar,
Phys. Rev. {\bf D56} (1997) 437.
\bibitem{yasuda}
O. Yasuda, TMUP-HEL-9603, hep-ph/9602342 (unpublished).
\bibitem{bugey}
B. Ackar et al., Nucl. Phys. {\bf B434}, (1995) 503.
\bibitem {kras}
G. S. Vidyakin et al., JETP Lett. {\bf 59}, 390 (1994).
\bibitem{msw}
S. P. Mikheyev and A. Smirnov, Nuovo Cim. {\bf 9C} (1986) 17; 
L. Wolfenstein, Phys. Rev. {\bf D17} (1978) 2369.
\bibitem{gnc}
Gargamelle Neutrino Collaboration, H. Deden et al.,
Nucl. Phys. {\bf B85} (1975) 269.
\bibitem{kajita1}
T. Kajita, {\it Physics and Astrophysics of Neutrinos} eds. M. 
Fukugita and A. Suzuki (Springer-Verlag, Tokyo, 1994), p559.
\bibitem{bc}
S. Baker and R.D. Cousin, Nucl. Instr. and Meth., {\bf 221} (1984) 437.
\bibitem{kajita2}T. Kajita, private communication.
\bibitem{nakahata}
M. Nakahata et. al., J. Phys. Soc. Japan, {\bf 55} (1986) 3786.
\bibitem{br}
P.R. Bevington and D.K. Robinson,
{\sl DATA REDUCTION AND ERROR ANALYSIS FOR THE PHYSICAL SCIENCES},
2nd ed. N.Y., McGraw-Hill, 1992.
\bibitem{solar}
See, e.g., J.N. Bahcall and R.K. Ulrich,
Rev. Mod. Phys. {\bf 60} (1988) 297;
J.N. Bahcall and M.H. Pinsonneault,
Rev. Mod. Phys. {\bf 64} (1992) 885;
J.N. Bahcall, R. Davis, Jr., P. Parker, A. Smirnov, R. Ulrich
eds., {\sl SOLAR NEUTRINOS: the first thirty years}
Reading, Mass., Addison-Wesley, 1994 and references therein.
\bibitem{yasuda2}
O. Yasuda in preparation.
\end{thebibliography}
\end{document}